\documentclass[10pt]{article}

\newcommand {\be}{\begin{equation}}
\newcommand {\ee}{\end{equation}}
\newcommand {\bea}{\begin{eqnarray}}
\newcommand {\eea}{\end{eqnarray}}

\newcommand {\vett}[1] {\mathbf{#1}}
\usepackage{a4}
\usepackage{psfig, graphicx}

\date{}

\title{Temperature-dependent density profiles of trapped boson-fermion
  mixtures}

\author{M. Amoruso*, A. Minguzzi*, S. Stringari$^\#$,
M. P. Tosi* and L. Vichi$^{\#}$\vspace{0.7cm} \\
{\it $^*$Istituto Nazionale di Fisica della Materia and Classe di
Scienze},\\{\it  Scuola Normale Superiore, Piazza dei Cavalieri
7,I-56126  Pisa, Italy}\\{\it and}\\
\it $^{\#}$Istituto Nazionale di Fisica della Materia and Dipartimento di
Fisica, \\ \it Universit\`{a} di Trento, I-38050 Povo, Italy \\}

\begin{document}
\maketitle

\vspace{2cm}

\begin{abstract}
 We present a semiclassical three-fluid model for a Bose-condensed
 mixture  of  interacting Bose and Fermi gases confined in harmonic
 traps at  finite temperature. The model is used to characterize the
 experimentally relevant behaviour of the equilibrium density profile
 of  the fermions with varying composition and temperature across the
 onset  of degeneracy, for coupling strengths relevant to a mixture of
 $^{39}$K and $^{40}$K atoms.
\end{abstract}

\vspace{5cm}


\newpage
\section*{}
        The achievement of Bose-Einstein condensation (BEC) in trapped
        gases of $^{87}$Rb [1], $^{23}$Na [2] and $^7$Li [3] has provided new
        impulse to the study of many-body and quantum statistical
        effects in dilute fluids at very low temperature. The
        formation of coexisting condensates by sympathetic cooling in
        a mixture of Rb atoms in two different internal states has
        also been achieved [4]. Trapping of fermionic species has been
        reported for $^6$Li [5] and $^{40}$K [6]. Trapped mixtures of bosonic
        and fermionic species are expected to become accessible to
        experiment in the near future.

        The density profiles of the separate fermionic and bosonic
        species in such mixtures can in principle be experimentally
        resolved. With this perspective we present in this work a
        semiclassical three-fluid model extending our earlier studies
        of the condensate fraction and internal energy of a trapped
        interacting Bose gas [7, 8] to a Bose-condensed mixture of
        interacting Bose and Fermi gases confined in a spherically
        symmetric trap at finite temperature. We assume that the
        mixture is in full thermal equilibrium: the efficiency of a
        Bose condensate in cooling slow impurity atoms has been
        discussed by Timmermans and C\^ot\'e [9]. The main emphasis
        of our calculations is on the evaluation of the density
        profile of the fermionic component with varying temperature
        and composition for values of the coupling strengths relevant
        to a mixture of $^{39}$K and $^{40}$K atoms. We also identify two
        parameters which govern the behaviour of the profile. In the
        limit of the Thomas-Fermi approximation at zero temperature we
        recover the equations used by M\o lmer [10] to describe the
        ground state of mixtures of bosons and fermions. Studies of
        the thermodynamic properties of ideal Fermi gases in harmonic
        traps have been reported by Butts and Rokhsar [11] and by
        Schneider and Wallis [12].

        Interaction effects are very small in the normal phase but
        become significant as condensation induces a density increase
        of the bosonic component near the centre of the trap. For
        values of the (repulsive) coupling strengths in the range of
        present interest, the condensate pushes both the bosonic
        noncondensate and the fermionic fluid towards the periphery of
        the trap. With regard to the latter fluid, such 'squeezing'
        drives an increase in its chemical potential. As we shall see
        below, these effects on the fermionic density profile are
        especially striking in boson-rich mixtures, but rather low
        temperatures relative to the BEC transition need to be
        reached. Of course, an increase of the fermion-boson coupling,
        as may be achieved by modifying the scattering lengths with
        external fields [13, 14] and by changing the parameters of the
        trap or the components of the mixture, will enhance the
        BEC-induced changes in the fermionic fluid.
 
        In the following we shall assume that the number of bosons in
        the trap is large enough that the kinetic energy term in the
        Gross-Pitaevskii equation for the wave function $\Psi(\vett
        r)$ of  the condensate can be neglected (see e.g. [15]). This
        corresponds to the so-called Thomas-Fermi limit and is in
        general a  good approximation except in the immediate
        neighbourhood  of the BEC transition temperature. It yields
        the  strong-coupling result
\be
\Psi^2(r)=[\mu_b-V_b^{ext}(r) - 2 g n_{nc}(r) - f n_f(r)]/g
\ee
when the quantity in the square bracket is positive, and $\Psi^2(r) =
0$ otherwise. Here, $g=4 \pi \hbar^2 a_b/m_b$ and $f=2 \pi
\hbar^2 a_f /m_r$ with $a_f$  and $a_b$  the boson-boson and
boson-fermion $s$-wave scattering lengths and $m_r=m_b m_f/(m_b+m_f)$
with $m_b$ and $m_f$  the atomic masses; $\mu_b(T)$  is the chemical
potential of the bosons at temperature $T$, $V_b^{ext}(r)=m_b
\omega_b^2r^2/2$ is a spherically symmetric external potential
confining the bosons, $n_{nc}(r)$  is the average distribution of
non-condensed bosons and $n_f(r)$ is that of the fermions. The factor
2 in the third term on the RHS of eqn (1) arises from exchange [16]
and we have neglected a term involving the off-diagonal density of
non-condensed bosons.

        As already proposed in early work on the confined Bose fluid
        [17 - 19], we treat both the non-condensed bosons and the
        fermions as ideal gases in effective potentials $V_b^{eff}(r)$
        and $V_f^{eff}(r)$  involving the relevant interactions. We write
\be
V_b^{eff}(r)=V_b^{ext}(r)+ 2g \Psi^2(r)+2 g n_{nc}(r) + fn_f(r)
\ee
and
\be
V_f^{eff}(r)=V_f^{ext}(r)+f \Psi^2(r) +f n_{nc}(r)\;,
\ee
with $V_f^{ext}(r)=m_f \omega_f^2 r^2/2$.  We are taking the fermionic
        component as a dilute, spin-polarized Fermi gas: the
        fermion-fermion interactions are then associated at leading
        order with $p$-wave scattering and are demonstrably negligible
        at the temperatures of present interest [11, 20]. We may then
        evaluate the thermal averages with standard Bose-Einstein and
        Fermi-Dirac distributions, taking the non-condensed particles
        to be in thermal equilibrium with the condensate at the same
        chemical potential $\mu_b(T)$  and the fermions at chemical
        potential  $\mu_f(T)$. In the semiclassical approximation we obtain
\be
n_f(r)=\frac{1}{h^3} \int d^3p \left \{ \exp \left[
        \left(\frac{p^2}{2m_f}+
        V_f^{eff}(r)-\mu_f\right)/k_BT\right]+1  \right\}^{-1}
\ee
and
\bea
n_{nc}(r)=\frac{1}{h^3}\int
        d^3p\left\{\exp\left[\left(\frac{p^2}{2m_b}+V_b^{eff}(r)-\mu_b
        \right)/k_BT\right]-1 \right\}^{-1}\nonumber \\=\left(\frac{2
        \pi  m_b k_BT}{h^2}\right)^{3/2}\sum_{j\geq 1}       
\frac{\exp[-j(V_b^{eff}(r)-\mu_b)/k_BT]}{j^{3/2}}\;.
\eea
The chemical potentials are determined from the total numbers of
        bosons and  fermions,
\be
N_b=\int d^3r\,[\Psi^2(r)+n_{nc}(r)]
\ee
and
\be
N_f=\int d^3 r\,n_f(r)\;.
\ee
These equations complete the self-consistent closure of the model.

        Before presenting relevant illustrative examples of the full
        numerical solution of the set of equations (1) - (7), it is
        useful to discuss a simplified form of the present three-fluid
        model (see also [7]). This is obtained by introducing an
        approximation which preserves only the repulsions exerted by
        the condensate: namely, we set to zero the last two terms in
        the RHS of eqns (1) and (2) and the last term in the RHS of
        eqn (3). Evidently, this approximation rests on the fact that
        both the non-condensed and the fermion component are very
        dilute gases and becomes all the more accurate as one moves
        towards boson-rich mixtures well below the BEC transition
        temperature. In this approximation we can (i) introduce a
        temperature-dependent scale length
        $R=(2\mu_b/m_b\omega_b^2)^{1/2}$ defining the radius outside
        which the condensate density vanishes, and (ii) scale the
        effective potential acting on the fermions by writing it in
        the form $V_f^{eff}(r)=\hbar \omega_f \tilde V_f^{eff}(x)$
        where  $x=r/R$ and
\be
\tilde V_f^{eff}(x)=
\left\{\begin{array}{cc}
        \frac{1}{2}\gamma \left[\lambda +(1-\lambda)x^2\right] & {\rm
        for  }  \hspace{0.3cm}x<1 \\
        \frac{1}{2}\gamma x^2  &{\rm for  } \hspace{0.3cm} x>1 \\
        \end{array}\right.
\ee
with
\be
\lambda=\frac{fm_b
        \omega_b^2}{g m_f \omega_f^2}
\ee
and
\be
\gamma=\frac{2\mu_bm_f\omega_f}{\hbar
        m_b \omega_b^2}=\left[15N_c a_b \left(\frac{m_b\omega_b} 
{\hbar}\right)^{1/2}\right]^{2/5}\frac{m_f\omega_f}{m_b\omega_b}\;.
\ee

Evidently, the dimensionless constant $\lambda$ controls the shape of
the effective potential seen by the fermions and hence their density
profile: it depends only on (i) the ratio of the two relevant
scattering  lengths and (ii) the ratio of the spring constants of the
two  traps. Instead the parameter $\gamma$, which depends on the
number  $N_c(T)$ of bosons in the condensate through their chemical
potential, controls the depth of the effective potential (in units of
$\hbar \omega_f$). It is easily seen from eqn (8) that for $\lambda >
1$ $\tilde V_f^{eff}(x)$ decreases from the value $\gamma\lambda/2$ at
the centre of the trap towards its minimum value $\gamma/2$ at $r=R$,
and increases thereafter because of the confinement. In this regime
the  fermions are squeezed away from the centre of the trap into a
shell  overlying both the outer part of the condensate and the
non-condensate. On the other hand, for $\lambda<1$ the minimum value
of  $\tilde V_f^{eff}(x)$ is $\gamma \lambda/2$ at the centre of the
trap:  namely, in this regime the condensate merely raises the bottom
of  the confining well for the fermions without changing the sign of
its  central curvature. 
        
The calculations that we report below refer to the fermion density
profiles  in a trap with $\omega_f=\omega_b= 100 \, {\rm s}^{-1}$. We
shall  first consider (left panels in Figures 1 and 2) the case
$a_b\simeq 80$ and $a_f\simeq 46$ Bohr radii for the boson-boson and
boson-fermion $s$-wave scattering lengths, corresponding to the
$^{39}$K-$^{40}$K mixture [21]. With the above values we find
$f\simeq 0.57 g$ and hence $\lambda <1$. We have also examined the
regime  $\lambda>1$ by assuming $f=2g$ (right panels in the
Figures). We have studied the dependence of the profiles on the
composition  of the mixture by considering two cases, namely (i)
$N_f=10^3$  and $N_b=10^6$ (top panels in the Figures) and (ii)
$N_f=N_b=10^4$ (bottom panels). The two characteristic temperatures of
the  mixture shown in the Figures are the BEC transition temperature
$k_BT_c=\hbar\omega_b (N_b/\zeta(3))^{1/3}$ for an harmonically
confined  ideal Bose gas and the Fermi temperature $T_f$. The latter
has been  calculated from the chemical potential of the fermions in
the  simplified model at zero temperature. The increase of $T_f$ due
to the  boson-fermion interactions, relative to the confined ideal-gas
value  $k_BT_f=\hbar \omega_f (6N_f)^{1/3}$, is quite large at
boson-rich  compositions (about 40$\%$ in Figure 1.a and 86$\%$ in Figure 1.c).

Figure
        1 compares our results for the density profile of the fermions
        at two  temperatures below $T_c$ with those for a confined
        Fermi gas  in the absence of all interactions. In this range
        of parameters the results obtained in the simplified model
        leading to  eqn (8) are practically indistinguishable from
        those  obtained from the full numerical solution of eqns (1) -
        (7).  The interactions induce major distortions of the fermion
        density profile near the centre of the trap in the
        $^{39}$K-$^{40}$K  mixture  at low temperatures and at
        boson-rich  compositions (see the curve in Figure 1.a
        referring to  $T/T_c = 0.1$ and $N_f=10^{-3}N_b$). The
        transition  from the first to the second regime of coupling
        strength  $\lambda$ is clearly shown by the comparison between
        the  left and right-hand panels in Figure 1. For boson-rich
        compositions at low temperature, the fermions are in fact
        almost  wholly expelled from the central region of the trap
        (see  Figure 1.c). This agrees with the results of the
        ground-state  calculations of M{\o }lmer [10]. The density
        profiles of  the bosons, that we do not show, are instead
        hardly  affected by the presence of the fermions in the range
        of  parameters that we have explored.

        It is also interesting to point out that for $T=0.1T_c$ the
        conditions of Figures 1.a and 1.c correspond to temperatures
        of the  same order as the Fermi temperature $T_f$. In this
        case the  effects of the interactions are more important than
        those  due to the quantum degeneracy of the Fermi gas. On the
        contrary, the conditions of Figures 1.b and 1.d at $T=0.1T_c$
        correspond to temperatures much smaller than $T_f$ and the
        effects of quantum degeneracy are much more important, as is
        illustrated by the comparison with the density profile for an
        ideal  Boltzmann gas at the same temperature (dots in Figures
        1.a and  1.b).

        Figure 2 shows the behaviour of the density $n_f(r=0)$ of
        fermions at  the centre of the trap as a function of
        temperature, for  the same coupling strengths and compositions
        as in  Figure 1. In Figure 2.a we have reported in an inset
        the  behaviour of the temperature derivative of $n_f(r=0)$, to
        display its upturn occurring as quantum degeneracy develops at
        sufficiently low values of $T/T_f$. Major effects should be
        expected  in boson-rich mixtures if the coupling strength can
        be  driven into the $\lambda>1$ regime, as is seen from Figure 2.c.

        In summary, we have studied the equilibrium density profile of
        a  fermionic fluid in a confined, Bose-condensed mixture of
        bosons  and fermions in dependence of composition and
        temperature.  The fermion density at the centre of the trap
        can be  used to detect the onset of degeneracy in the Fermi
        gas with  decreasing temperature. Although we have assumed
        spherically  symmetric traps, the model can easily be extended
        to  asymmetric confinements through a simple change of variables.

        As a final remark we notice that the density profile that we
        have  calculated for the fermionic component immediately
        yields a  semiclassical density of states $\rho_f(E)$ for the
        calculation of thermodynamic properties through the relation
\be
\rho_f(E)=2\pi
(2m_f/h^2)^{3/2}\int
d^3r
[E-V_f^{eff}(r)]^{1/2}
\;.
\ee
A similar relation holds, of course, for the density of states of the
        non-condensate [17-19, 7]. In the simplified model leading to
        eqn (8),  that we have seen to be quite accurate compared with
        a fully  self-consistent treatment of the mixture in the range
        of  parameters that we have explored, the integral in eqn (11)
        is  easily evaluated analytically. Calculations of
        thermodynamic  properties for boson-fermion mixtures will be
        reported  elsewhere.

\subsection*{Acknowledgments}

We are very grateful to J. P. Burke and C. Greene for providing us
with values  of the scattering lengths of K isotopes prior to
publication.  Useful discussions with M. L. Chiofalo, S. Giorgini and
G. M.  Tino are also acknowledged. This work is supported by the
Istituto  Nazionale di Fisica della Materia through the Advanced
Research  Project on BEC.

\newpage
\subsection*{Figure captions}

\begin{itemize}
\item{}
 Figure
 1.
 Fermion density profile $n_f(r)$ (in units of $a_0^{-3}$, with
 $a_0=(\hbar/m_b \omega)^{1/2}$ and $\omega$ the characteristic
 frequency of  the harmonic trap) {\it versus} distance $r$ from the
 centre of  the trap (in units of $a_0$), for two values of the
 reduced  temperature $T/T_c$ ( $T/T_c = 0.1$ and 0.5). The dashed
 curves and  the dots are for an ideal Fermi gas and for an ideal
 Boltzmann  gas (at $T/T_c = 0.1$) in the absence of bosons,
 respectively. The  full curves are the results of the numerical
 solution of  the full set of eqns (1) - (7). The values of the
 parameters for  the various panels are as follows: (a) $f=0.57 g$,
 $N_f =  10^3$ and $N_b = 10^6$; (b) $f=0.57 g$ and $N_f=N_b  = 10^4$;
 (c)  $f=2g$, $N_f = 10^3$ and $N_b = 10^6$; (d) $f=2g$ and $N_f = N_b
 =  10^4$. The values of the BEC transition temperature $T_c$ and of
 the  Fermi temperature $T_f$ are also indicated.

\item{}
Figure
2.
Fermion
density
$n_f(0)$
at
the
centre
of
the
trap
(in
units
of
$a_0^{-3}$)
{\it
versus}
reduced temperature $k_BT/\hbar\omega$. The results of the simplified
model in  eqn (8) (long-dashed curves) are compared with those for an
ideal  Fermi gas (short-dashed curves). The arrows mark the BEC
transition  temperature and the Fermi temperature of the mixture. The
inset in  panel (a) shows the temperature derivative of $n_f(0)$ (in
reduced  units) versus reduced temperature.

\end{itemize}
\begin{figure}
\begin{center}
\includegraphics[scale=0.78]{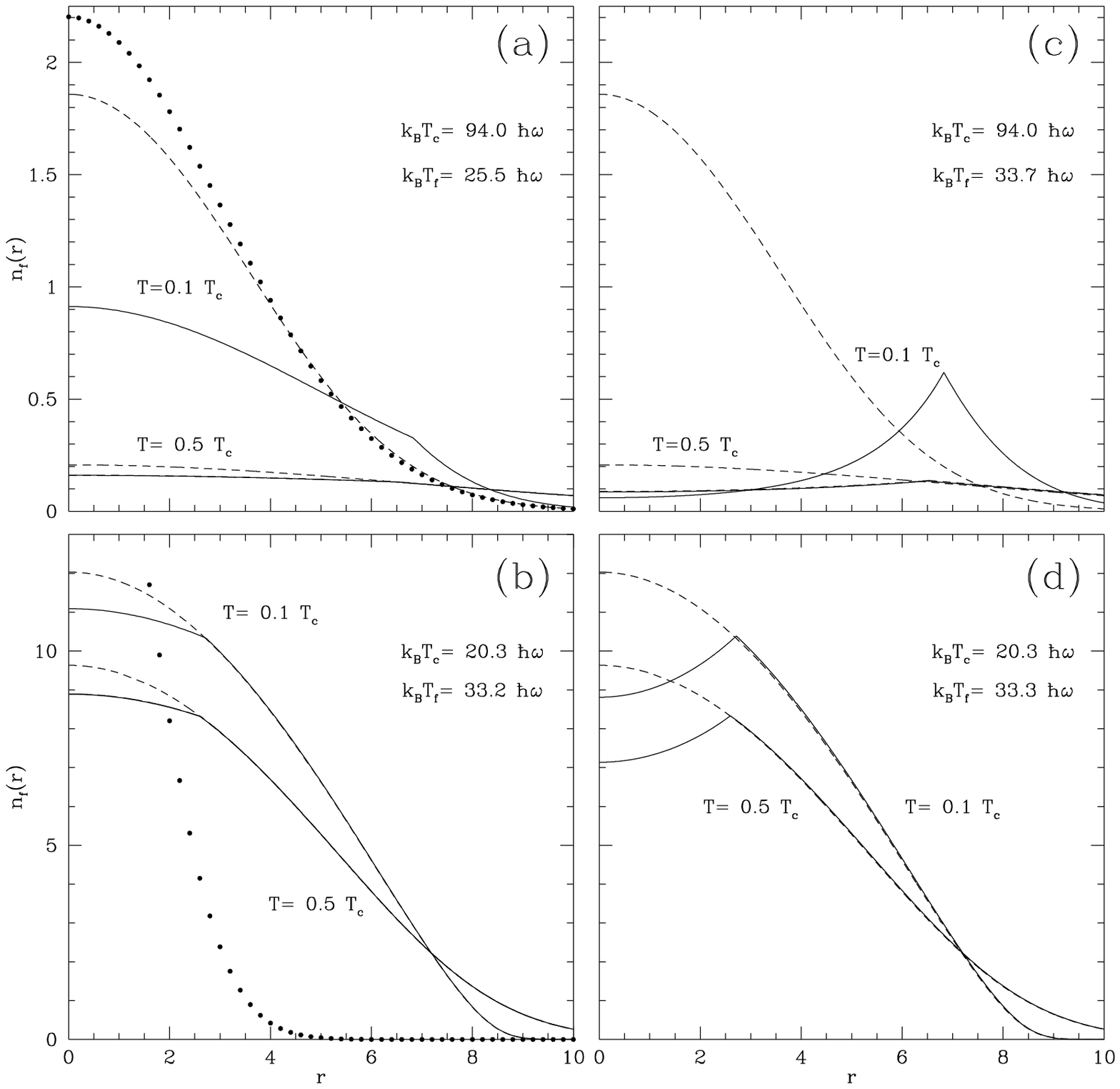} 
\end{center}
\end{figure}
\begin{figure}
\begin{center}
\includegraphics[scale=0.78]{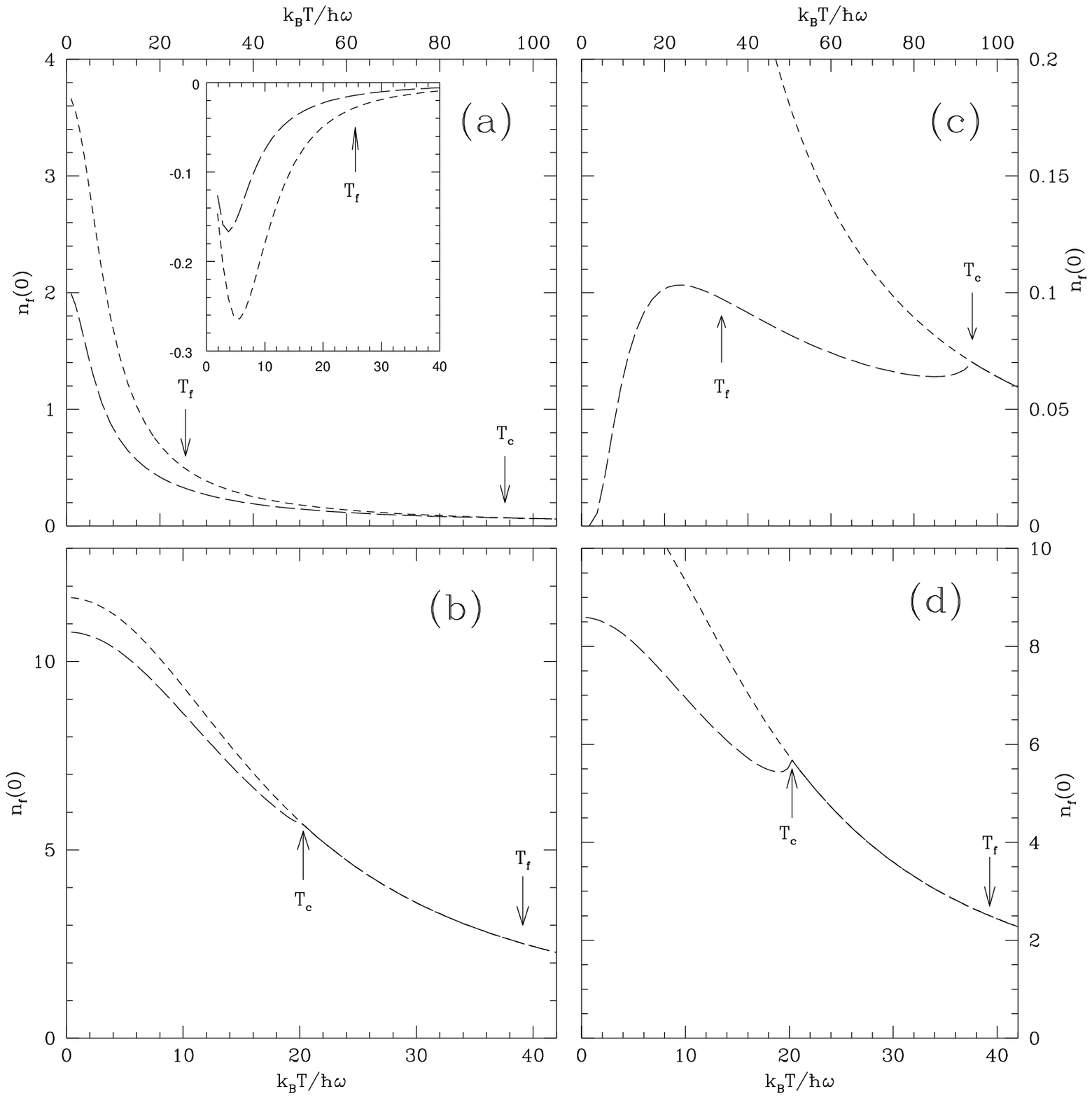} 
\end{center}
\end{figure}


\begin{thebibliography}{10}
\bibitem{1}
M. H. Anderson, J. R. Ensher, M. R. Matthews, C. E. Wieman and
E. A. Cornell,  Science 269, 198 (1995).
\bibitem{2} K. B. Davis, M.-O. Mewes, M. R. Andrews, N. J. van Druten,
  D. S.  Durfee, D. M. Kurn and W. Ketterle, Phys. Rev. Lett. 75, 3969 (1995).
\bibitem{3} C. C. Bradley, C. A. Sackett, J. J. Tollett and
  R. G. Hulet, Phys.  Rev. Lett. 75, 1687       (1995) and 79, 1170 (1997).
\bibitem{4}  
  C. J. Myatt, E. A. Burt, R. W. Ghrist, E. A. Cornell and
  C. E. Wieman, Phys.  Rev.     Lett. 78, 586 (1997).
\bibitem{5} 
  W. I. McAlexander, E. R. I. Abraham, N. W. M. Ritchie,
  C. J. Williams, H. T.  C. Stoof       and R. G. Hulet, Phys. Rev. A
  51, R871  (1995).
\bibitem{6}  
  F. S. Cataliotti, E. A. Cornell, C. Fort, M. Inguscio, F. Marin, M.
  Prevedelli, L. Ricci and      G. M. Tino (in press).
\bibitem{7}
  A. Minguzzi, S. Conti and M. P. Tosi, J. Phys.: Condens. Matter 9,
  L33 (1997).
\bibitem{8}  
  S. Giorgini, L. P. Pitaevskii and S. Stringari, Phys. Rev. Lett. 78,
  3987  (1997).
 \bibitem{9}
  E. Timmermans and R. C\^ot\'e, Phys. Rev. Lett. 80, 3419 (1998).
\bibitem{10}
K.
M{\o
}lmer,
Phys.
Rev.
Lett.
80,
1804
(1998).
\bibitem{11}    D. A. Butts and D. S. Rokhsar, Phys. Rev. A 55, 4346 (1997).
\bibitem{12}
J.
Schneider
and
H.
Wallis,
Phys.
Rev.
A 57,
1253
(1998).
\bibitem{13}
P.
O. Fedichev, Yu. Kagan, G. V. Shlyapnikov and J. T. M. Walraven,
Phys. Rev.  Lett.       77, 2913 (1996).
\bibitem{14}
S.
Inouye, M. R. Andrews, J. Stenger, H.-J. Miesner, D. M. Stamper-Kurn
and W.          Ketterle, Nature 392, 151 (1998).
\bibitem{15}
S.
Giorgini,
L.
P.
Pitaevskii
and
S.
Stringari,
Phys.
Rev.
A 54,
R4633
(1996).
\bibitem{16}
A.
Griffin,
Phys.
Rev.
B 53,
9341
(1996).
\bibitem{17}    V. V. Goldman, I. F. Silvera and A. J. Leggett,
  Phys. Rev. B  24, 2870 (1981).
\bibitem{18}
D.
A.
Huse
and
E.
D.
Siggia,
J.
Low
Temper.
Phys.
46,
137
(1982).
\bibitem{19}    V. Bagnato, D. E. Pritchard and D. Kleppner,
  Phys. Rev. A 35,  4354 (1987).
\bibitem{20}
For
the
specific
case
of
$^{40}$K
in a
mixture
with
$^{39}$K,
taking
a
mean
interparticle
distance
$r_0\simeq
10 $
nm
and a
de
Broglie
wavelength
$\lambda_{dB}$
corresponding
to
$k_BT\simeq
100
\hbar
\omega_f$
with
$\omega_f\simeq
100\,{\rm
s}^{-1}$, we    estimate an upper limit $r_0^6/(\lambda_{dB}^4a_f^2)$
for the  ratio between the cross sections for $p$-wave fermion-fermion
scattering and for $s$-wave fermion-boson scattering as being of order
$10^{-4}$.
\bibitem{21}    J. P. Burke and C. Greene, private communication (March 1998).
\end{thebibliography}
\end{document}